\documentclass[conference]{IEEEtran}
\IEEEoverridecommandlockouts
\usepackage{siunitx}
\usepackage{cite}
\usepackage{amsmath,amssymb,amsfonts}
\usepackage{graphicx}
\usepackage{epstopdf}
\usepackage{textcomp}
\usepackage{xcolor}
\usepackage{cite}
\usepackage{amsmath,amssymb,amsfonts}
\usepackage{textcomp}
\usepackage{xcolor}
\usepackage[numbers]{natbib}
\usepackage{graphicx}
\usepackage{amsmath,amssymb,amsfonts}
\usepackage{booktabs}
\usepackage{graphicx}
\usepackage{textcomp}
\usepackage{comment}
\usepackage{float}
\usepackage{verbatim}
\usepackage{pgfplots}
\usepackage{caption}
\usepackage[caption=false,font=normalsize,labelfont=sf,textfont=sf]{subfig}
\usepackage{url}
\usepackage{multirow}
\usepackage{booktabs}
\usepackage{hhline}
\usepackage{tikz}
\usepackage{hyperref}
\usepackage{mdframed}
\usepackage[caption=false,font=normalsize,labelfont=sf,textfont=sf]{subfig}
\usepackage{caption}
\usepackage[utf8]{inputenc}
\usepackage{algorithm}
\usepackage{placeins}
\usepackage{placeins}

\usepackage{algpseudocode}
\usepackage{algorithm}

\def\BibTeX{{\rm B\kern-.05em{\sc i\kern-.025em b}\kern-.08em
    T\kern-.1667em\lower.7ex\hbox{E}\kern-.125emX}}

\def\BibTeX{{\rm B\kern-.05em{\sc i\kern-.025em b}\kern-.08em
    T\kern-.1667em\lower.7ex\hbox{E}\kern-.125emX}}
\begin{document}

\title{

Emotion Recognition with Minimal Wearable Sensing: Multi-domain Feature, Hybrid Feature Selection, and Personalized vs. Generalized Ensemble Model Analysis \\
{\footnotesize \textsuperscript{*}Note: Accepted in IEEE BIBM 2025}
\thanks{Acknowledgment: Funded by the European Union (AI4HOPE, 101136769)}
}

\author{
\IEEEauthorblockN{
Muhammad Irfan\IEEEauthorrefmark{1}\IEEEauthorrefmark{6}\IEEEauthorrefmark{2},
Anum Nawaz\IEEEauthorrefmark{2}\IEEEauthorrefmark{4},
Ayse Kosal Bulbul \IEEEauthorrefmark{5},
Riku Kl\'en\IEEEauthorrefmark{1},
Abdulhamit Subasi\IEEEauthorrefmark{5} \IEEEauthorrefmark{3},
Tomi Westerlund\IEEEauthorrefmark{2},
\\ 
Wei Chen\IEEEauthorrefmark{6}
}
\IEEEauthorblockA{
\IEEEauthorrefmark{1}Institute of Biomedicine, Turku University Hospital, University of Turku, Finland, and University of Sydney, Australia 
\\
\IEEEauthorrefmark{2}Turku Intelligent Embedded and Robotic Systems lab, University of Turku, Finland
\\
\IEEEauthorrefmark{3} Department of Computer Science, Effat University, Jeddah, Saudi Arabia
\\
\IEEEauthorrefmark{4} School of Information Science and Technology, Fudan University, China\\
\IEEEauthorrefmark{5} Institute of Biomedicine, University of Turku, Finland  
\\
\IEEEauthorrefmark{6} School of Biomedical Engineering, University of Sydney, Australia.}
\{muirfa, anum.nawa, akbulb,
tomi.westerlund, abdulhamit.subasi, 
riku.klen\}@utu.fi\\

nanum18@fudan.edu.cn, wei.chenbme@sydney.edu.au
}

\maketitle

\begin{abstract}


Negative emotions are linked to the onset of neurodegenerative diseases and dementia, yet they are often difficult to detect through observation. Physiological signals from wearable devices offer a promising noninvasive method for continuous emotion monitoring. In this study, we propose a lightweight, resource-efficient machine learning approach for binary emotion classification, distinguishing between negative (sadness, disgust, anger) and positive (amusement, tenderness, gratitude) affective states using only electrocardiography (ECG) signals. The method is designed for deployment in resource-constrained systems, such as Internet of Things (IoT) devices, by reducing battery consumption and cloud data transmission through the avoidance of computationally expensive multimodal inputs. We utilized ECG data from 218 CSV files extracted from four studies in the Psychophysiology of Positive and Negative Emotions (POPANE) dataset, which comprises recordings from 1,157 healthy participants across seven studies. Each file represents a unique subject emotion, and the ECG signals, recorded at \SI{1000}{\hertz}, were segmented into 10-second epochs to reflect real-world usage. Our approach integrates multidomain feature extraction, selective feature fusion, and a voting classifier. We evaluated it using a participant-exclusive generalized model and a participant-inclusive personalized model. The personalized model achieved the best performance, with an average accuracy of 95.59\%, outperforming the generalized model, which reached 69.92\% accuracy. Comparisons with other studies on the POPANE and similar datasets show that our approach consistently outperforms existing methods. This work highlights the effectiveness of personalized models in emotion recognition and their suitability for wearable applications that require accurate, low-power, and real-time emotion tracking. Code availability at GitHub...

\end{abstract}


\section{Introduction}
\label{sec:intro}

Emotions are not just fleeting experiences; they are deeply intertwined with brain function, physical health, and cognitive aging~\cite{eslinger2021neuroscience}. In the context of neurodegenerative diseases like Alzheimer's, emotional states play a pivotal role not only in symptom manifestation but also in disease progression. Emerging research indicates that persistent negative emotional patterns, such as repetitive negative thinking (RNT), are associated with increased deposition of amyloid and tau proteins in the brain, hallmarks of Alzheimer's disease. Individuals exhibiting higher levels of RNT experienced greater cognitive decline and had more significant amyloid and tau accumulation, suggesting that chronic negative emotions may be a preventable risk factor for dementia~\cite{schlosser2020repetitive}.

Furthermore, behavioral and psychological symptoms of dementia (BPSD), including depression, apathy, agitation, delusions, and anxiety, affect up to 90\% of those diagnosed with dementia during their illness~\cite{devshi2015prevalence}. These symptoms often intensify over time and are strongly associated with increased rates of institutionalization. Moreover, individuals recently diagnosed with dementia commonly experience a range of emotions, including grief, anger, fear, and disbelief. Understanding and addressing these emotional responses are essential for providing empathetic, effective care~\cite{devshi2015prevalence}.

Emotions play a unique role in shaping mental health outcomes. Emotions like gratitude, amusement, and tenderness activate reward pathways in the brain and release neurotransmitters like serotonin and dopamine. These neurochemical changes enhance cognitive flexibility, improve problem-solving, and build psychological resilience. The broaden-and-build theory, for example, suggests that positive emotions lead to an expansion of an individual's repertoire of thoughts and actions, leading to the development of lasting personal resources~\cite{yang2024emotion}.

Conversely, negative emotions such as sadness, anger, and disgust are linked to increased activity in brain regions like the amygdala and insula, which process threat and emotional pain~\cite{clevelandclinic-amygdala}. While short-term activation of these regions can be adaptive, chronic exposure to negative emotional states has been shown to elevate stress hormones, impair cognitive function, and increase vulnerability to mental health disorders~\cite{cobb2024emotions}.

\subsubsection*{\textbf{Challenges}}

To enable real-time emotion recognition, wearable sensors have become a crucial component in both health monitoring and human-computer interaction, particularly for individuals with dementia. These devices offer non-invasive, continuous tracking of physiological states, which is essential for understanding emotional well-being in populations with impaired self-reporting abilities. However, one of the primary challenges in wearable technology lies in power consumption and data transmission. For instance, high-frequency, multi-stream data collection can overwhelm Bluetooth bandwidth, leading to unreliable data transfer to cloud servers~\cite{polar-ble-sdk}. This limitation has been acknowledged in the latest wearable technologies, where initiating multiple high-frequency streams simultaneously may cause data loss due to excessive communication traffic. Additionally, rapid battery drainage remains a significant issue, as noted in studies involving commercial devices such as the Polar 360~\cite{polar-ble-sdk}.

\subsubsection*{\textbf{Research Aims}}

This study investigates the comparative performance of personalized versus generalized machine learning models in a binary emotion classification task (distinguishing between negative and positive emotional states) using biosignal data collected from wearable devices. The goal is to determine whether personalized solutions offer significant advantages in accuracy and responsiveness for real-time emotion monitoring. Furthermore, we address the limitations of current wearable emotion recognition systems by adopting a minimal sensing approach, specifically utilizing a single physiological signal. Our methodology emphasizes advanced feature selection and multidomain feature extraction to maximize information efficiency while maintaining a low computational footprint. We evaluate a suite of traditional machine learning algorithms that are both lightweight and interpretable, thereby supporting real-world deployment scenarios. This consideration is particularly critical in applications involving individuals with cognitive decline or mental health challenges, where system accessibility, simplicity, and unobtrusiveness are essential.

\subsubsection*{\textbf{Main Contributions of the Article}}

\begin{enumerate}

    \item We developed a robust pipeline for classifying positive or negative emotions using electrocardiography (ECG) signals, with a focus on minimal sensing to respond to practical constraints in wearable health monitoring, particularly for dementia patients.
    
    \item We developed a novel two-stage feature engineering framework combining multidomain feature extraction, including time-domain, frequency-domain, and R-R interval–based heart rate variability features, with a hybrid feature selection methodology. 
    
    \item We developed an ensemble model (EnsemNet) and
    performed a comparative evaluation of 19 machine learning models, culminating in ensemble strategies such as voting and stacking classifiers to enhance predictive performance.
    
    \item We investigated the trade-offs between personalized and generalized learning paradigms using participant-exclusive and participant-inclusive splits to assess generalizability in real-world deployment scenarios.
    
    \item All experiments are conducted on the Psychophysiology of Positive and Negative Emotions (POPANE) dataset, a multimodal biosignal dataset from 1157 participants, with a specific focus on ECG data, to enable low-power, real-time emotion recognition.

\end{enumerate}

The remainder of this paper is organized as follows: \textbf{Section~\ref{sec:rw}} reviews existing research on emotion classification using various biosignals, as well as audio and video data. \textbf{Section~\ref{sec:method}} details the datasets used, preprocessing techniques, hybrid feature selection and fusion strategies, and the implementation of the proposed methodology. \textbf{Section~\ref{sec:abl}} provides the ablation analysis of hybrid feature selection and ensemble models. \textbf{Section~\ref{sec:results}} presents the experimental results, compares the proposed approach with existing methods, and evaluates its performance under various scenarios. Finally, \textbf{Section~\ref{sec:conclusion}} summarizes the main findings and discusses the broader implications of this work.

\section{Related Work}
\label{sec:rw}

Emotion recognition using physiological signals has gained significant traction in recent years, particularly due to the increasing availability of wearable sensors and advances in machine learning. Signals such as electrodermal activity (EDA), ECG, electroencephalography (EEG), and photoplethysmography (PPG) are known to capture various physiological properties such as sleep~\cite{irfan2025smart,irfan2025novel,irfan2025multidomain,nahliis2023ensemble}, autonomic nervous system responses linked to emotional states \cite{10918671,yang2024emotion,10950432,rainville2006basic}. Unlike biomedical image data~\cite{irfan2024evaluation} or behavioral data such as facial expressions or speech, biosignals offer a noninvasive, continuous, and relatively privacy-preserving means of monitoring emotional well-being.

With the proliferation of consumer-grade wearable devices, like Fitbit and Apple Watch, that now support real-time heart rate, ECG, and even EDA sensing, there is a growing opportunity to apply emotion recognition models outside laboratory environments. These models hold particular promise for personalized mental health support, especially in populations that struggle with self-reporting, such as individuals with dementia or autism spectrum disorders.

Much of the prior work in this space has focused on emotion classification using multimodal datasets such as Wearable Stress and Affect Detection (WESAD), a Database for Emotion Recognition through EEG and ECG Signals (DREAMER), and a multimodal database for implicit personality and Affect Recognition using commercial physiological sensors (ASCERTAIN)~\cite{gahlan2024aflemp,kanjo2019deep,panahi2021application}. While multimodal inputs can improve classification accuracy, they also pose significant challenges for real-world deployment on wearables, where battery life, computational resources, and wireless data transmission via Bluetooth Low Energy (BLE) are constrained. Reducing the number of required input signals without sacrificing performance is thus a critical research goal.

Additionally, many emotion recognition systems rely on audio or video data, which raises concerns about user privacy and social acceptability. Biosignal-based systems, especially those using a single, unobtrusive sensor, offer a more discreet alternative~\cite{alasiry2025efficient,noroozi2017audio}.

Another persistent challenge in emotion recognition from biosignals is the generalizability of machine learning models. Cross-subject validation, where models are trained on data from one group of participants and tested on a separate group, offers a more realistic evaluation framework compared to random splits. However, this approach often reveals significant performance degradation, primarily due to high inter-subject variability in physiological responses to emotion.

In wearable applications, this variability becomes especially critical. Since wearables typically serve individual users, personalized models, trained on a user’s physiological data, can be much more effective. These models tend to outperform generalized ones by capturing unique, subject-specific signal patterns, which generalized models may fail to model accurately. This has led to growing interest in personalized or semi-personalized approaches that adapt to individual physiology~\cite{schmidt2018introducing,sah2021stress,li2024comparison}. Generalized models benefit from diverse training data but often underperform in real-world use. Personalized models, despite needing individual calibration, offer greater accuracy and reliability in mobile and health-monitoring applications~\cite{10842226,li2024comparison}.

In this work, we contribute to this line of research by evaluating emotion recognition models trained exclusively on ECG data. Our aim is twofold: to minimize sensor input for practical deployment and to explore how model personalization affects performance. Using the POPANE dataset, one of the largest publicly available biosignal datasets, with recordings from 1,157 participants across seven studies~\cite{behnke2022psychophysiology}, we compare two modeling strategies: (1) a semi-personalized model and (2) a generalized model trained with participant-inclusive and participant-exclusive data splits.

Our findings indicate that the semi-personalized approach consistently outperforms generalized alternatives, especially in distinguishing between negative (sadness, disgust, anger) and positive (amusement, tenderness, gratitude) emotional states. These results are validated on the POPANE dataset, the largest publicly available biosignal dataset, underscoring the potential of lightweight, ECG-based emotion recognition models for real-world use.

\section{Materials and Methods}
\label{sec:method}

\subsection{Study Flowchart}

\begin{figure*}[th]
    \centering
    \includegraphics[width=0.9\linewidth]{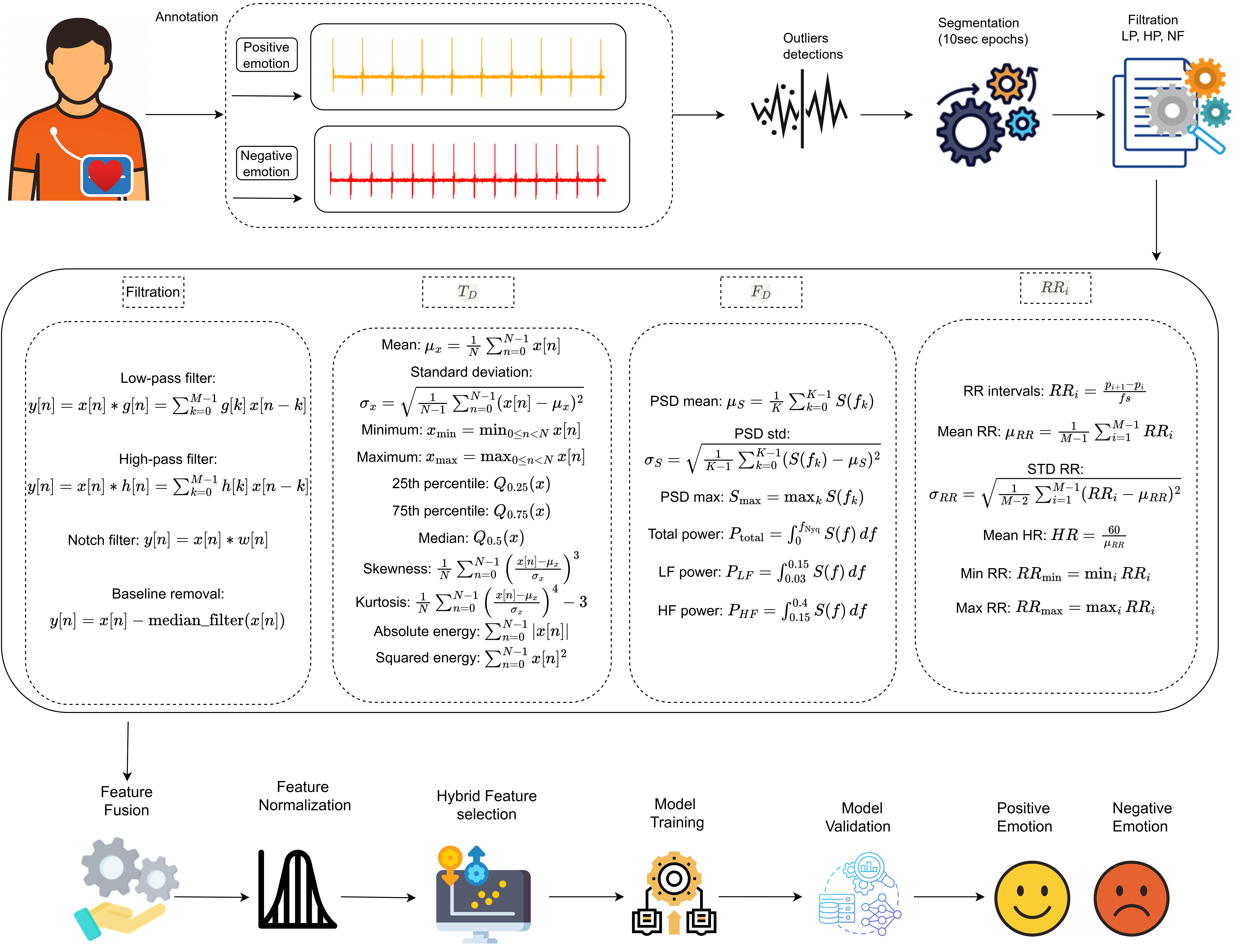}
    \caption{An Overview of the Study}
    \label{fig:study-overview}
\end{figure*}

Fig.~\ref{fig:study-overview} illustrates the key steps of the study, including data filtration, multi-domain feature extraction, feature fusion, hybrid feature selection, model training, validation, and classification. To prevent data leakage, feature normalization and selection functions were fitted exclusively on the training data, with the corresponding transformations subsequently applied to the test data. 

\subsection{Dataset}

This study utilizes the publicly available Psychophysiology of POPANE dataset, which offers a large-scale collection of psychophysiological responses associated with emotional experiences. The POPANE dataset includes recordings from 1,157 healthy young adults, gathered across seven independent studies. Participants’ affective states and physiological responses were continuously recorded during both resting baselines and emotionally evocative conditions. Emotion elicitation methods included film clips, emotional imagery (pictures), speech preparation tasks, and expressive writing, covering a diverse spectrum of both positive and negative emotions such as amusement, anger, disgust, excitement, fear, gratitude, sadness, tenderness, and threat.

Physiological data were collected using ECG, impedance cardiography (ICG), EDA, PPG (used for blood volume pulse), respiratory sensors, and skin temperature sensors. In total, the dataset comprises approximately 725 hours of multimodal recordings. To date, the POPANE database represents the most extensive and publicly accessible resource for investigating the psychophysiology of emotional experiences. However, in this study, we aim to use a single modality (ECG) as our focus is to reduce battery consumption, lower data transfer to the cloud, and analyze the results of ECG for positive vs. negative emotions.

\subsection{Data Preparation}

To ensure a clean ECG signal suitable for emotion recognition, we implemented a multi-stage preprocessing pipeline. The raw signals were first filtered using a notch filter to eliminate powerline interference at 50 \(Hz\), employing an infinite impulse response (IIR) notch filter with a quality factor of 30. Following this, we applied a high-pass Butterworth filter (cutoff frequency = 0.01 \(Hz\), order = 4) to remove slow baseline drifts and a low-pass Butterworth filter (cutoff frequency = 40 \(Hz\), order = 3) to suppress high-frequency noise. In addition, we applied a baseline correction procedure by subtracting a low-frequency component (obtained via a low-pass filter with a 0.05 \(Hz\) cutoff) from the signal, effectively isolating dynamic signal fluctuations associated with physiological responses.

After filtering and baseline correction, the preprocessed signals were segmented into fixed-length epochs of 10 seconds (non-overlapping). This segmentation approach supports real-world applications, such as real-time monitoring and classification of emotional states, by enabling emotion recognition within short and manageable time windows.

In the personalized solution, we perform a within-subject split for binary classification by dividing each subject-emotion group (with $\geq$12 epochs) into 75\% training and 25\% testing, ensuring the model learns from and evaluates on the same group of subjects. In the generalized solution, we perform a leave-one-group-out split by randomly selecting 20\% of subjects for testing and using the remaining 80\% for training. This ensures the model is evaluated on unseen subjects, testing its ability to generalize.

\subsection{Multidomain Feature Extraction}

To effectively capture the complex physiological dynamics associated with emotional responses, we extracted features from multiple domains: time, frequency, and heart rate variability (HRV). Multidomain feature extraction enables a richer representation of physiological signals by leveraging complementary information across temporal, spectral, and cardiovascular dimensions.

For the ECG signal epoch \( x = (x_1, x_2, \ldots, x_N) \), sampled at frequency \( f_s = 1000 \,\text{Hz} \), the following features were computed:

\subsubsection*{Time-Domain Features}

Time-domain features summarize the statistical and morphological properties of the signal:

\begin{itemize}
    \item Mean: \( \mu = \frac{1}{N} \sum_{i=1}^N x_i \), where \( N \) is the number of data points.

    \item Standard Deviation: \( \sigma = \sqrt{\frac{1}{N} \sum_{i=1}^N (x_i - \mu)^2} \)
    \item Minimum and Maximum: \( \min(x), \max(x) \)
    \item 25th and 75th Percentiles: \( Q_1, Q_3 \)
    \item Median: \( \text{median}(x) \)
    \item Skewness: 
    \[
    \text{Skew}(x) = \frac{1}{N} \sum_{i=1}^N \left( \frac{x_i - \mu}{\sigma} \right)^3
    \]
    \item Kurtosis: 
    \[
    \text{Kurtosis}(x) = \frac{1}{N} \sum_{i=1}^N \left( \frac{x_i - \mu}{\sigma} \right)^4
    \]
    \item Absolute Sum: \( \sum_{i=1}^N |x_i| \)
    \item Energy: \( \sum_{i=1}^N x_i^2 \)
\end{itemize}

\subsubsection*{Frequency-Domain Features}

We computed Power Spectral Density (PSD) using Welch’s method and derived the following features:

\begin{itemize}
    \item Mean, standard deviation, and maximum of PSD
    \item Total power: 
    \[
    P_{\text{total}} = \int P(f) \, df
    \] Where $f$ is frequency. 
    \item Low-Frequency (LF) Power (0.03--0.15 \(Hz\))
    \item High-Frequency (HF) Power (0.15--0.4 \(Hz\))
\end{itemize}

LF and HF powers are physiologically relevant markers of autonomic nervous system activity.

\subsubsection*{Heart Rate Variability (HRV) Features}

Peak detection was applied to estimate RR intervals \( RR_i \), defined as:

\[
RR_i = \frac{t_{i+1} - t_i}{f_s}
\]  $f_s$ is sample frequency 

From these intervals, we computed:

\begin{itemize}
    \item Mean and standard deviation of RR intervals
    \item Heart Rate (HR): \( HR = \frac{60}{\text{mean}(RR)} \)
    \item Minimum and maximum RR intervals
\end{itemize}
 
Segments with insufficient peaks were zero-padded to preserve feature dimensionality.

\subsubsection*{Label Assignment and Data Structuring}

Each segment was labeled according to an emotion class mapped from raw numeric labels. Extracted features and corresponding labels were compiled for use in downstream classification tasks.

\subsection{Hybrid Feature Selection}

To optimize the feature set for classification, we employed a two-stage hybrid feature selection approach. First, an embedded method using an \textit{Extra Trees Classifier} (500 estimators) was applied to the scaled training data to rank features by importance. The top 18 features were retained based on the feature importance scores.

In the second stage, a univariate filter method, \textit{SelectKBest} with ANOVA F-test (\texttt{f\_classif}), was used to further select the top \( k = 13 \) features from the previously reduced set. This combination of embedded and filter methods allowed us to capture both interaction-based importance and statistical relevance to the target labels.

\subsection{Classifiers}

To evaluate the classification performance of the extracted features, we implemented a diverse set of machine learning classifiers spanning multiple algorithmic families. Linear models included logistic regression with both \texttt{liblinear} and \texttt{newton-cg} solvers, as well as stochastic gradient descent (SGD) classifiers using \texttt{log\_loss} and \texttt{hinge} losses. Naive Bayes models included GaussianNB and BernoulliNB variants. For Support Vector Machines (SVM), we tested radial basis function (RBF), linear, and polynomial kernels, all wrapped in a pipeline with standard scaling.

We also employed tree-based models, including decision trees with maximum depths of 5 and 10 and random forest classifiers with 300 estimators and depths of 10 and 20. Extra Trees classifiers (XTRA) were configured with \texttt{max\_features} set to \texttt{sqrt} and \texttt{log2}. Additionally, we used Linear Discriminant Analysis (LDA) as a baseline linear classifier.

Instance-based models were represented by K-Nearest Neighbors (KNN), with configurations of 5 and 7 neighbors using uniform and distance-based weighting, respectively. Neural networks were explored using multilayer perceptrons (MLP) with one hidden layer of 100 neurons and a two-layer architecture with 50 neurons each. Ensemble methods included gradient boosting (200 estimators, learning rate 0.05) and AdaBoost (100 estimators, learning rate 0.5), as well as bagging with 100 base estimators and sampling ratios of 0.8 for both samples and features.

In addition to these individual classifiers, we also experimented with ensemble meta-models, including a voting classifier (both hard and soft voting strategies) and a stacking classifier. These ensemble approaches were designed to leverage the strengths of multiple base models, thereby improving classification accuracy and robustness. The most effective strategy is EnsemNet, and its pseudocode is given in Algorithm~\ref{alg:EnsemNet}. The algorithm initializes several base classifiers and constructs two ensemble models, ``Voting\_Soft," which combines five diverse learners using soft voting, and ``FinalModel" (EnsemNet), a higher-level ensemble that includes key base models and ``Voting\_Soft"; each model is trained on the training data, evaluated on the test data, and its performance metrics are recorded, enabling a comprehensive comparison across all models.

\begin{algorithm}
\caption{Ensemble Model: EnsemNet}
\label{alg:EnsemNet}
\begin{algorithmic}[1]
\State {Define base models:}
\State \quad \texttt{LR\_Lib} $\leftarrow$ LogisticRegression with \texttt{liblinear}, scaled
\State \quad \texttt{NB\_Gaussian} $\leftarrow$ GaussianNB
\State \quad \texttt{SVM\_Poly} $\leftarrow$ SVM (poly kernel), scaled
\State \quad \texttt{MLP\_50\_50} $\leftarrow$ MLP with two hidden layers (50, 50), and scaled
\State \quad \texttt{GB} $\leftarrow$ GradientBoostingClassifier
\State \quad \texttt{Ada} $\leftarrow$ AdaBoostClassifier

\State {Define Voting\_Soft model:}
\State \quad \texttt{Voting\_Soft} $\leftarrow$ VotingClassifier(soft voting) with:
\State \quad \quad [\texttt{NB\_Gaussian, SVM\_Poly, MLP\_50\_50, GB, Ada}]

\State {Define FinalModel:}
\State \quad \texttt{FinalModel} $\leftarrow$ VotingClassifier (soft voting) with:
\State \quad \quad [\texttt{NB\_Gaussian, MLP\_50\_50, SVM\_Poly, Voting\_Soft}]

\State {Add FinalModel and all other models to dictionary} \texttt{models}

\ForAll{\texttt{(name, model)} in \texttt{models}}
    \State Train model: \texttt{model.fit($X_{\text{train}}$, $y_{\text{train}}$)}
    \State Predict labels: $\hat{y} \leftarrow \texttt{model.predict}(X_{\text{test}})$

    \State Compute Accuracy: \texttt{acc} $\leftarrow$ \texttt{AccuracyScore($y_{\text{test}}, \hat{y}$)}

\EndFor

\end{algorithmic}
\end{algorithm}

\section{Ablation Analysis}
\label{sec:abl}

\subsection{Model Performance for Generalized Solutions: With vs. Without Hybrid FS}

We conducted an ablation study to evaluate the individual and combined contributions of feature selection methods, specifically, KBest (univariate selection) and a hybrid approach combining ExtraTrees with KBest. Table~\ref{tab:ablation_kbest} reports the classification accuracy across ten models under three scenarios: no feature selection (No FS), KBest only, and the hybrid method (Hybrid FS).

The results show that while some models (e.g., SVM, Ensemble, Voting\_Soft) benefit significantly from hybrid feature selection, others (e.g., RF, KNN) perform better with KBest alone. Notably, the ensemble model shows the highest gain in accuracy when using the hybrid method, suggesting that the combination of embedded and filter-based selection can enhance model generalizability.

\begin{table}[th]
\caption{Ablation analysis: Performance impact of single feature selection (KBest only) vs. hybrid feature selection (ExtraTrees + KBest)}
\label{tab:ablation_kbest}
\centering
\begin{tabular}{@{}p{0.2\linewidth} p{0.15\linewidth} p{0.18\linewidth} p{0.18\linewidth} p{0.12\linewidth}@{}}
\toprule
\textbf{Model} & \textbf{Accuracy (No FS)} & \textbf{Accuracy (Hybrid FS)} & \textbf{Accuracy (KBest only)} & \textbf{Best FS} \\
\midrule

LR        & 59.20   &  56.00    & 57.31     & No FS \\

GNB        & 69.40   & 69.11   &  67.47  & No FS \\

SVM        & 62.00    &  65.24    &  65.65   & KBest \\

DT        & 60.00    & 60.16    &  55.48   & Hybrid \\

RF        & 58.80   &  56.30    & 62.60    & KBest \\

XTRA       & 52.40   & 52.80    & 58.53    & KBest \\

KNN        & 54.50   & 54.26    & 57.72    & KBest \\

MLP        & 63.80   & 62.60    & 63.00  & No FS \\

Voting\_Soft & 63.20   &  68.70   & 65.44    & Hybrid \\

Ensemble   & 64.80   &  \textbf{69.92}   & 66.05    & Hybrid \\
\bottomrule
\end{tabular}
\end{table}

\subsection{Model Performance for Personalized Solutions: With Hybrid FS vs. SelectKBest vs. Without FS}

To evaluate how different feature selection strategies influence model performance in personalized solutions, we compared results across three configurations:

\begin{enumerate}
    \item Without feature selection (No FS)
    \item With hybrid feature selection (Hybrid FS)
    \item With SelectKBest feature selection
\end{enumerate}

\begin{table}[th]
\centering
\caption{Top model accuracies across feature selection strategies}
\begin{tabular}{@{}p{0.2\linewidth} p{0.15\linewidth} p{0.18\linewidth} p{0.18\linewidth} p{0.12\linewidth}@{}}
\toprule
\textbf{Model} & \textbf{Accuracy (No FS)} & \textbf{Accuracy (Hybrid FS} & \textbf{Accuracy (KBest only)} & \textbf{Best FS} \\
\midrule
XTRA     & 91.90 & 93.27 & 89.91 & Hybrid \\
Ensemble & 91.13 & 95.60 & 88.07 & Hybrid \\
Bagging       & 90.06 & 90.67 & 88.69 & Hybrid \\
RF       & 89.76 & 90.52 & 88.07 & Hybrid \\
MLP      & 88.84 & 88.99 & 87.16 & Hybrid \\
KNN      & 87.46 & 88.69 & 81.19 & Hybrid \\
\hline
\end{tabular}
\end{table}

\subsubsection*{Observations}

Hybrid feature selection consistently produced the best results. The final model, using a combination of ensemble learners, achieved the highest accuracy of 0.952 with hybrid FS. This confirms the benefit of using more sophisticated feature selection techniques, especially when combined with powerful ensemble models. Ensemble-based methods such as ExtraTrees, Bagging, and Random Forest showed strong performance even without feature selection but still improved when hybrid FS was applied. In contrast, SelectKBest led to moderate improvements over no FS but did not match the performance gains seen with hybrid FS. Simpler classifiers like SGD performed poorly across all configurations and did not benefit significantly from feature selection.

\section{Experimental Results and Discussion}
\label{sec:results}

\subsection{Comparison of ML Models: Personalized vs. Generalized Models}

\begin{table}[th]
\caption{Performance of ML models in Personalize solution}
\label{tab:personalized}
\centering
\begin{tabular}{@{}p{0.32\linewidth} p{0.12\linewidth} p{0.12\linewidth} p{0.12\linewidth} p{0.12\linewidth}@{}}
\toprule
\textbf{Model} & \textbf{Accuracy } & \textbf{Precision} & \textbf{Recall} & \textbf{F1} \\
\midrule
{Ensemble}                 & {95.59} & {95.71} & {95.62} & {95.59} \\
XTRA\_Sqrt                     & 93.27 & 93.27 & 93.28 & 93.27 \\
\midrule
XTRA\_Log2                     & 93.27 & 93.27 & 93.28 & 93.27 \\
Bagging                              & 90.67 & 90.71 & 90.69 & 90.67 \\
\midrule
RF\_Depth20                & 90.52 & 90.52 & 90.52 & 90.52 \\
MLP\_50\_50                           & 88.99 & 89.18 & 89.02 & 88.98 \\ \midrule
KNN\_7\_Distance                      & 88.69 & 88.79 & 88.71 & 88.68 \\
KNN\_5\_Uniform                       & 88.69 & 88.91 & 88.72 & 88.67 \\\midrule
MLP\_100                              & 87.92 & 87.97 & 87.94 & 87.92 \\
RF\_Depth10                & 87.77 & 88.06 & 87.81 & 87.75 \\
\bottomrule
\end{tabular}
\end{table}

The results presented in Table~\ref{tab:personalized} and Table~\ref{tab:genl} demonstrate that personalized models significantly outperform generalized models across all evaluation metrics. The best-performing personalized model (an ensemble approach) achieved 95.59\% accuracy, compared to 69.92\% for the best generalized ensemble. This performance gap is consistently reflected in precision, recall, and F1 score, highlighting the superior effectiveness of tailoring models to individual data distributions.

In particular, classifiers such as ExtraTrees, RandomForest, and MLP show strong improvements when used in a personalized setup, suggesting that models benefit from user-specific patterns rather than relying on a generalized approximation. These findings support the hypothesis that personalized machine learning offers more reliable and accurate predictions, especially in domains with user-dependent behavior or preferences.

\begin{table}[th]
\caption{Performance of ML models in Generalized solution}
\label{tab:genl}
\centering
\begin{tabular}{@{}p{0.32\linewidth} p{0.12\linewidth} p{0.12\linewidth} p{0.12\linewidth} p{0.12\linewidth}@{}}
\toprule
\textbf{Model} & \textbf{Accuracy } & \textbf{Precision } & \textbf{Recall} & \textbf{F1} \\
\midrule
{Ensemble}                & {69.92} & {70.26} & {69.73} & {69.65} \\
NaiveBayes\_Gaussian               & 69.11 & 70.41 & 68.77 & 68.35 \\ \midrule
Voting\_Soft                       & 68.70 & 68.78 & 68.58 & 68.57 \\
SVM\_Poly                          & 65.24 & 73.62 & 64.49 & 61.19 \\\midrule
SVM\_Linear                        & 64.02 & 66.51 & 63.51 & 62.10 \\
MLP\_50\_50                        & 63.41 & 63.39 & 63.35 & 63.35 \\ \midrule
MLP\_100                           & 62.60 & 62.58 & 62.52 & 62.51 \\ 
Voting\_Hard                       & 63.82 & 64.05 & 63.61 & 63.44 \\ \midrule
SVM\_RBF                           & 60.98 & 61.07 & 60.78 & 60.63 \\ \
GradientBoosting                   & 60.57 & 61.39 & 60.84 & 60.19 \\
\bottomrule
\end{tabular}
\end{table}

\subsection{Discussion}
\label{sec:discussion}

This study set out to examine whether a single physiological signal (ECG) could be used effectively to distinguish between positive and negative emotional states. Our findings demonstrate that with careful feature engineering and model selection, ECG alone can provide a strong signal for emotion recognition, especially in personalized contexts. The significant accuracy gap between personalized and generalized models highlights the importance of user-specific physiological patterns. Personalized models consistently achieved better results, with the best model reaching an accuracy of 95.59\%, in contrast to 69.92\% for the best generalized model. This difference highlights the challenge of inter-subject variability in emotional responses, a persistent issue in physiological computing, and confirms the need for adaptive or user-calibrated systems in real-world applications. The hybrid feature selection strategy, merging the strengths of embedded (ExtraTrees) and univariate (SelectKBest) methods, played a critical role in improving model performance. Models that integrated this selection pipeline not only performed better but also remained computationally efficient, making them suitable for deployment in wearable devices with limited processing power. KNN classifiers, particularly those using distance-based weighting, also performed well in personalized settings. Their success suggests that ECG features exhibit localized structures that can be effectively exploited using proximity-based learning. This reinforces the utility of instance-based models in scenarios where personalization is possible and training data is user-specific.

However, a key trade-off of personalized models is their dependence on user-specific training data, which may not always be feasible at the start of system use. Future approaches could investigate transfer learning or adaptive calibration techniques to mitigate this limitation. 

\section{Conclusion}
\label{sec:conclusion}

This work introduced a simple and scalable approach to emotion recognition, tailored for real-time wearable applications. By focusing on a single input modality, ECG, we can reduce system complexity while still achieving competitive performance. Through a combination of multidomain feature extraction, hybrid feature selection, and lightweight ensemble modeling, our framework demonstrated strong predictive capability. The personalized model achieved a peak accuracy of 95.59\%, significantly outperforming its generalized counterpart. These findings validate the potential of ECG as a standalone modality for emotion recognition, especially when models are adapted to individual users. Importantly, this approach supports the broader goals of low-power, privacy-preserving, and continuous emotional monitoring, attributes essential for deployment in health-focused wearable systems. Populations such as individuals with cognitive impairments stand to benefit significantly from such technologies, where minimal sensing and unobtrusive monitoring are critical.

Future research directions include exploring semi-supervised personalization, real-world deployment on embedded hardware, and incorporating contextual signals such as activity or time-of-day to further refine emotion prediction models. By addressing these aspects, we aim to move closer to scalable and inclusive emotional health support systems powered by everyday wearable devices.

\bibliographystyle{ieeetr}
\bibliography{irfan}

\begin{thebibliography}{10}

\bibitem{eslinger2021neuroscience}
P.~J. Eslinger, S.~Anders, T.~Ballarini, S.~Boutros, S.~Krach, A.~V. Mayer, J.~Moll, T.~L. Newton, M.~L. Schroeter, R.~de~Oliveira-Souza, {\em et~al.}, ``The neuroscience of social feelings: mechanisms of adaptive social functioning,'' {\em Neuroscience \& Biobehavioral Reviews}, 2021.

\bibitem{schlosser2020repetitive}
M.~Schlosser, H.~Demnitz-King, T.~Whitfield, M.~Wirth, and N.~L. Marchant, ``Repetitive negative thinking is associated with subjective cognitive decline in older adults: a cross-sectional study,'' {\em BMC psychiatry}, vol.~20, pp.~1--10, 2020.

\bibitem{devshi2015prevalence}
R.~Devshi, S.~Shaw, J.~Elliott-King, E.~Hogervorst, A.~Hiremath, L.~Velayudhan, S.~Kumar, S.~Baillon, and S.~Bandelow, ``Prevalence of behavioural and psychological symptoms of dementia in individuals with learning disabilities,'' {\em Diagnostics}, vol.~5, no.~4, pp.~564--576, 2015.

\bibitem{yang2024emotion}
X.~Yang, H.~Yan, A.~Zhang, P.~Xu, S.~H. Pan, M.~I. Vai, and Y.~Gao, ``Emotion recognition based on multimodal physiological signals using spiking feed-forward neural networks,'' {\em Biomedical Signal Processing and Control}, vol.~91, p.~105921, 2024.

\bibitem{clevelandclinic-amygdala}
{Cleveland Clinic}, ``Amygdala: What it is and what it controls,'' 2023.
\newblock Accessed: 2025-05-23.

\bibitem{cobb2024emotions}
C.~C. (Amey), ``Emotions, aggression, and stress.'' \url{https://pressbooks.library.vcu.edu/psyc629/chapter/emotions-aggression-stress/}, 2024.
\newblock Accessed: 2025-05-23.

\bibitem{polar-ble-sdk}
{Polar Electro Oy}, ``{Polar BLE SDK}.'' \url{https://github.com/polarofficial/polar-ble-sdk}, 2025.
\newblock Accessed: 2025-05-23.

\bibitem{irfan2025smart}
M.~Irfan, L.~Wang, Y.~Xu, A.~Subasi, C.~Chen, R.~Klen, T.~Westurlund, and W.~Chen, ``Smart iot-based solutions for neonatal sleep stratification: Single-dual channel eeg, adaptiselect, multview fusion, \& rotational ensemble stacking,'' {\em IEEE Internet of Things Journal}, 2025.

\bibitem{irfan2025novel}
M.~Irfan, A.~Subasi, Z.~Tang, L.~Wang, Y.~Xu, C.~Chen, T.~Westurlund, and W.~Chen, ``A novel nicu sleep state stratification: Multiperspective features, adaptive feature selection and ensemble model,'' {\em IEEE Transactions on Biomedical Engineering}, 2025.

\bibitem{irfan2025multidomain}
M.~Irfan, L.~Wang, H.~Shahid, Y.~Xu, A.~Subasi, A.~Munawar, N.~Mustafa, C.~Chen, T.~Westurlund, and W.~Chen, ``Multidomain selective feature fusion and stacking based ensemble framework for eeg-based neonatal sleep stratification,'' {\em IEEE Journal of Biomedical and Health Informatics}, 2025.

\bibitem{nahliis2023ensemble}
A.~NAHLIIS, C.~CHEN, Y.~XU, L.~WANG, and A.~NAWAZ, ``An ensemble voting approach with innovative multi-domain feature fusion for neonatal sleep stratification,'' 2023.

\bibitem{10918671}
M.~Irfan, A.~Subasi, Z.~Tang, L.~Wang, Y.~Xu, C.~Chen, T.~Westurlund, and W.~Chen, ``A novel nicu sleep state stratification: Multiperspective features, adaptive feature selection and ensemble model,'' {\em IEEE Transactions on Biomedical Engineering}, pp.~1--13, 2025.

\bibitem{10950432}
M.~Irfan, L.~Wang, Y.~Xu, A.~Subasi, C.~Chen, R.~Klen, T.~Westerlund, and W.~Chen, ``Smart iot-based solutions for neonatal sleep stratification: Single-dual channel eeg, adaptiselect, multview fusion, \& rotational ensemble stacking,'' {\em IEEE Internet of Things Journal}, pp.~1--1, 2025.

\bibitem{rainville2006basic}
P.~Rainville, A.~Bechara, N.~Naqvi, and A.~R. Damasio, ``Basic emotions are associated with distinct patterns of cardiorespiratory activity,'' {\em International journal of psychophysiology}, vol.~61, no.~1, pp.~5--18, 2006.

\bibitem{irfan2024evaluation}
M.~Irfan, A.~Subasi, N.~Mustafa, T.~Westerlund, and W.~Chen, ``An evaluation of pretrained convolutional neural networks for stroke classification from brain ct images,'' in {\em Applications of Artificial Intelligence in Healthcare and Biomedicine}, pp.~111--135, Elsevier, 2024.

\bibitem{gahlan2024aflemp}
N.~Gahlan and D.~Sethia, ``Aflemp: Attention-based federated learning for emotion recognition using multi-modal physiological data,'' {\em Biomedical Signal Processing and Control}, vol.~94, p.~106353, 2024.

\bibitem{kanjo2019deep}
E.~Kanjo, E.~M. Younis, and C.~S. Ang, ``Deep learning analysis of mobile physiological, environmental and location sensor data for emotion detection,'' {\em Information Fusion}, vol.~49, pp.~46--56, 2019.

\bibitem{panahi2021application}
F.~Panahi, S.~Rashidi, and A.~Sheikhani, ``Application of fractional fourier transform in feature extraction from electrocardiogram and galvanic skin response for emotion recognition,'' {\em Biomedical Signal Processing and Control}, vol.~69, p.~102863, 2021.

\bibitem{alasiry2025efficient}
A.~Alasiry, M.~Al-Hussain, M.~Turki-Hadj~Alouane, and N.~Ben Hadj-Alouane, ``Efficient audio-visual emotion recognition approach,'' {\em Multimedia Tools and Applications}, pp.~1--25, 2025.

\bibitem{noroozi2017audio}
F.~Noroozi, M.~Marjanovic, A.~Njegus, S.~Escalera, and G.~Anbarjafari, ``Audio-visual emotion recognition in video clips,'' {\em IEEE Transactions on Affective Computing}, vol.~10, no.~1, pp.~60--75, 2017.

\bibitem{schmidt2018introducing}
P.~Schmidt, A.~Reiss, R.~Duerichen, C.~Marberger, and K.~Van~Laerhoven, ``Introducing wesad, a multimodal dataset for wearable stress and affect detection,'' in {\em Proceedings of the 20th ACM international conference on multimodal interaction}, pp.~400--408, 2018.

\bibitem{sah2021stress}
R.~K. Sah and H.~Ghasemzadeh, ``Stress classification and personalization: Getting the most out of the least,'' {\em arXiv preprint arXiv:2107.05666}, 2021.

\bibitem{li2024comparison}
J.~Li, P.~Washington, {\em et~al.}, ``A comparison of personalized and generalized approaches to emotion recognition using consumer wearable devices: Machine learning study,'' {\em JMIR AI}, vol.~3, no.~1, p.~e52171, 2024.

\bibitem{10842226}
M.~Irfan, L.~Wang, H.~Shahid, Y.~Xu, A.~Subasi, A.~Munawar, N.~Mustafa, C.~Chen, T.~Westurlund, and W.~Chen, ``Multidomain selective feature fusion and stacking based ensemble framework for eeg-based neonatal sleep stratification,'' {\em IEEE Journal of Biomedical and Health Informatics}, pp.~1--10, 2025.

\bibitem{behnke2022psychophysiology}
M.~Behnke, M.~Buchwald, A.~Bykowski, S.~Kupi{\'n}ski, and L.~D. Kaczmarek, ``Psychophysiology of positive and negative emotions, dataset of 1157 cases and 8 biosignals,'' {\em Scientific Data}, vol.~9, no.~1, p.~10, 2022.

\end{thebibliography}

\end{document}